\documentclass{article}

\usepackage{amsmath,amssymb,amsfonts,dcolumn,color,graphicx,graphics,latexsym,placeins,epsfig}
\usepackage{graphicx}  
\usepackage{amsmath}   
\usepackage{amssymb}   
\usepackage{bm} 
\usepackage{dcolumn}
\usepackage{color}
\usepackage{footmisc}
\usepackage{mathrsfs}
\usepackage{amsfonts}
\usepackage{varioref}
\usepackage{slashed}
\usepackage{footnote}
\usepackage[belowskip=-17pt,aboveskip=15pt]{caption}
\usepackage{amsmath}
\usepackage{makeidx}
\usepackage{amssymb}
\usepackage{graphicx}
\usepackage{bmpsize}

\usepackage[margin=1.5in]{geometry}
\usepackage{graphicx}
\usepackage{dcolumn}
\usepackage{amsmath,amssymb,mathtools}
\usepackage{epstopdf}
\usepackage{verbatim}
\usepackage[colorlinks=true, citecolor=blue, urlcolor = blue, linkcolor= blue, 
bookmarks=true]{hyperref}
\def\e{\epsilon}
\def\p{\partial}

\def\be{\begin{equation}}
\def\ee{\end{equation}}

\def\e{\epsilon}
\def\p{\partial}
\labelformat{section}{Section #1} 
\labelformat{subsection}{Section #1} 
\labelformat{subsubsection}{Section #1}
\labelformat{subsubsubsection}{Section #1}
\labelformat{equation}{Eq.~(#1)} 
\labelformat{figure}{Fig.~#1} 
\labelformat{subfigure}{Fig.~\thefigure#1} 
\labelformat{table}{Tab.~#1} 
\labelformat{appendix}{Appendix #1}
\vspace{25pt}
\title{\bf Stationary black holes and stars in the Brans-Dicke theory with $\Lambda >0$ revisited }
\author{$^1$Md Sabir Ali\footnote{alimd.sabir3@gmail.com},\;\,\,$^{2,3}$Sourav Bhattacharya\footnote{sbhatta.physics@jadavpuruniversity.in; On lien from IIT Ropar, Punjab, India},\, and\, $^3$Shagun Kaushal\footnote{2018phz0006@iitrpr.ac.in}\\
\small{$^1$Department of Physical Sciences, Indian Institute of Science Education and Research Kolkata, Mohanpur, WB741246, India}\\
\small{$^2$Relativity and Cosmology Research Centre, Department of Physics, Jadavpur University, Kolkata 700 032, India}\\
\small{$^3$Department of Physics, Indian Institute of Technology Ropar, Rupnagar, Punjab 140 001, India}\\}

\begin{document}

\maketitle

\begin{abstract}
\noindent
It was shown a few years back  that for a stationary regular black hole or star solution in the Brans-Dicke theory with a positive cosmological constant $\Lambda$, endowed with a de Sitter or cosmological event horizon in the asymptotic region, not only there exists no non-trivial field configurations, but also the inverse Brans-Dicke parameter $\omega^{-1}$ must be vanishing. This essentially reduces the theory to Einstein's General Relativity.  The assumption of the existence of the cosmological horizon was crucial for this proof. However, since the Brans-Dicke field $\phi$, couples directly to the $\Lambda$-term in the energy-momentum tensor as well as $\Lambda$ acts as a source in $\phi$'s equation of motion, it seems reasonable to ask : can $\phi$ become strong instead and screen the effect of $\Lambda$, at very large scales, so that the asymptotic de Sitter structure is replaced by some alternative, yet still acceptable   boundary condition?  In this work we analytically argue that  no such alternative exists, as long as the spacetime  is assumed to be free of any naked curvature singularity.  We further support this result by providing explicit numerical computations. Thus we conclude that in the presence of a positive $\Lambda$, irrespective of whether the asymptotic de Sitter boundary condition is imposed or not, a regular stationary  black hole or  even a star solution in the Brans-Dicke theory always necessitates $\omega^{-1}=0$,  and thereby reducing the theory to General Relativity. The qualitative differences of this result with that of the standard no hair theorems are  also pointed out.
\end{abstract}
  
\bigskip
\section{Introduction}\label{S1}
The classical black hole no hair theorem~\cite{Bekenstein:1971hc} states that there exists no non-trivial field configuration at the exterior of a stationary black hole spacetime, save the long range gauge fields,   related to the uniqueness of stationary black hole spacetimes, 
see~\cite{Bekenstein:1998aw}  for a vast review and list of references. The no hair theorems, their violations, and associated uniqueness properties  are much well studied, see~\cite{Herdeiro:2015waa} and references therein. Most of these efforts have been devoted to the asymptotically flat spacetimes. However, the overwhelming observational evidences of accelerated expansion of our current universe suggests that there is a strong possibility that it is endowed with some form of the dark energy, an exotic matter field with negative isotropic pressure. A positive cosmological constant $\Lambda$ is the simplest and phenomenologically very successful model of the same~\cite{Weinberg:2008zzc}. What happens to the black hole no hair or uniqueness properties in the presence of a positive $\Lambda$? An important feature of such spacetimes, in addition to the black hole horizon,  is certainly the existence of a de Sitter or  cosmological event horizon as the outer causal boundary, thereby making the asymptotic structure very different compared to  that of $\Lambda \leq 0$, e.g.~\cite{Gibbons:1977mu}. Can this horizon bring into non-trivial boundary effects? We refer our reader to~\cite{Chambers:1994sz, Bhattacharya:2007ap, Bhattacharya:2011dq, Bhattacharya:2013hvm} and references therein for discussion on black hole no hair theorems and their violations in such scenario.

Despite their phenomenological successes, the actual nature of dark energy and dark matter remain elusive so far. This has lead the community to plunge into research in various  gravity and dark energy theories alternative to Einstein's General Relativity in recent times, see e.g.~\cite{Clifton:2011jh} for a vast review and references therein. Such alternative models  mimic the dark sector chiefly via some dynamical matter fields or modification of the Einstein-Hilbert action.  The Brans-Dicke theory in particular, is the  prototype of the scalar-tensor class of theories~\cite{Brans:1961sx, Brans:1962zz},
\begin{eqnarray}
S= \int \sqrt{-g}\, d^4 x \left[\phi R - 2\Lambda - \frac{\omega}{\phi} (\nabla \phi)^2 +{\cal L}_M\right]
\label{bd1}
\end{eqnarray}
where the scalar $\phi$ is the Brans-Dicke field, whose inverse acts as a local and dynamical gravitational `constant' and ${\cal L}_M$ stands collectively for the matter Lagrangian density.  $\omega$ is  the Brans-Dicke parameter and as $\omega \to \infty$, the theory reduces to Einstein's General Relativity. We shall set $\phi\, (\omega^{-1}=0) = 1$. In other words, we shall work in a unit where $16\pi G =1$. 

The proof of no hair theorem for \ref{bd1} in the asymptotically flat spacetime can be seen in~\cite{Hawking}. However, one can have non-trivial configurations for $\phi$ in asymptotically flat non-black hole spacetimes like the sun~\cite{Brans:1961sx}. We further refer our reader to e.g.~\cite{Sotiriou:2011dz}-\cite{Nojiri:2020blr} and references therein for various aspects of black holes and large scale structures in the Brans-Dicke and some other viable  alternative gravity models. The extension of the no Brans-Dicke hair theorem of~\cite{Hawking} with a positive $\Lambda$ was considered a few years back in~\cite{Bhattacharya:2015iha}, with an asymptotic de Sitter boundary condition. It was shown that  not only any non-trivial field configuration for $\phi$ is excluded, but also the existence of the cosmological horizon reduces the theory to Einstein's General Relativity (i.e., $\omega^{-1}=0$).  Moreover, similar conclusion was shown to exist for a stationary star spacetime. Being a theory getting constrained, clearly these results are in stark contrast to {\it any} existing no hair theorems, which predict {\it only} about the field configurations.

 Now, even though the asymptotic de Sitter boundary condition seems to be reasonable, as we argue in \ref{S2}, perhaps it cannot be unique, chiefly owing to the fact that $\Lambda$ acts as an omnipresent source to the Brans-Dicke field via a Poisson equation, \ref{bd2}. Can we have hairy black hole and star solutions with some alternative asymptotic structure? Or at least, is it possible to just have the field configuration constrained as of~\cite{Hawking}, and leave $\omega$ unaffected? The answers to both these questions are negative, as no such non-singular alternative asymptotic structure exists, shown below in \ref{S2}, \ref{S3} and \ref{S4}, both analytically and numerically. In order to prove this, we fix the boundary conditions on the black hole event horizon, as described and argued in~\ref{S2}. 
 Accordingly we conclude that, in the presence of a positive $\Lambda$ and irrespective of whether the asymptotic de Sitter boundary condition is assumed to hold or not, a stationary black hole or star solution in the Brans-Dicke theory essentially necessitates $\omega^{-1}=0$, thereby reducing the theory to General Relativity, as long as there is no naked curvature singularity in the spacetime.  

\section{Non-existence of black holes  with generic asymptotic condition}\label{S2}

\noindent
 The equations of motion corresponding to \ref{bd1} are given by
\begin{eqnarray}
&&R_{\mu\nu}= \frac{\Lambda (2\omega +1)}{\phi (2\omega +3)}g_{\mu\nu}+\frac{T_{\mu\nu}}{\phi}- \frac{T (\omega +1)}{\phi (2\omega +3)}g_{\mu\nu}+\frac{\omega}{\phi^2} (\nabla_{\mu}\phi) (\nabla_{\nu}\phi)+ \frac{\nabla_{\mu}\nabla_{\nu} \phi}{\phi}\nonumber\\
&&\Box \phi = \frac{T-4\Lambda}{2\omega +3} \qquad (\omega \neq -3/2)
\label{bd2}
\end{eqnarray}
 where $T_{\mu\nu}$ is the energy-momentum tensor corresponding to ${\cal L}_M$ and $T$ is its trace.  From \ref{bd2}, the Ricci scalar is found to be
\begin{eqnarray}
R= \frac{ 2\omega (4\Lambda-T) }{\phi (2\omega +3)}+\frac{\omega}{\phi^2}(\nabla_{\mu}\phi)(\nabla^{\mu}\phi)
\label{bd3}
\end{eqnarray}

Let us look for  regular stationary black hole solutions admitted by \ref{bd1}. We  take the ansatz for a static and spherically symmetric metric  
\be
ds^2= -f(r) dt^2 + h(r)dr^2 + r^2 \left(d\theta^2 +\sin^2 \theta d\varphi^2 \right)
\label{bd4}
\ee  
Since the geometry is static and spherically symmetric, we shall take, by the virtue of \ref{bd2}, that $\phi$ is explicitly independent of time and is a function of the radial coordinate only. From the staticity of $\phi$, the equation of motion for $\phi$ becomes
\be
D_{\mu}\left(\sqrt{f} D^{\mu} \phi \right) =   -\frac{\sqrt{f}(4\Lambda-T)}{2\omega +3} 
\label{bd5}
\ee  
where $D_{\mu}$ is the spacelike covariant derivative on the $(r,\theta,\phi)$-hypersurface of \ref{bd4}.   Setting $T_{\mu\nu}=0$, and  multiplying the above equation by $e^{\e \phi}$ ($\e=\pm 1$), and integrating it by parts on that spacelike hypersurface  (say, $\Sigma$),
  we have
\be
\int_{\p \Sigma} \sqrt{\frac{f}{h}} e^{\e \phi}\,  \partial_r \phi =   \int_{\Sigma}\sqrt{f} e^{\e \phi}\left[\e(D_{\mu}\phi)(D^{\mu}\phi)  -\frac{4\Lambda}{2\omega +3}\right]
\label{bd6}
\ee  
where the left hand side consists of the boundary integrals over 2-spheres. One of the boundaries is the black hole event horizon where $f,\,h^{-1}=0$. If in addition we assume  the existence of a cosmological event horizon  as a boundary  in the asymptotic region, we have $f,\,h^{-1}=0$ there as well. The regularity of the field and its derivative on the horizon, as it turns out to be necessary for  non-naked singular  horizons then implies that the boundary integrals vanish. The derivative term appearing on the right hand side of \ref{bd6}, being spacelike, is positive definite. We then set $\e =-1\, (\e=+1)$ for $2\omega+3 >0\,(2\omega+3<0)$, yielding not only a constant $\phi$, but also $\omega^{-1}=0$. This essentially rules out the theory, provided a cosmological event horizon, in addition to the black hole event horizon exists, as was shown in~\cite{Bhattacharya:2015iha}. Similar result was shown to 
exist for stationary star solutions as well. 
  
Thus the existence of a cosmological event horizon plays a crucial role above. Such existence  seems to be plausible, as intuitively it seems  that in the presence of positive $\Lambda$'s repulsive effects, the Brans-Dicke field will become very diluted at  large scales. This assumption is strengthened by the solar system constraint showing the weakness of the Brans-Dicke parameter, $|\omega| \gtrsim 40\,000$, e.g.~\cite{Clifton:2011jh}. However, the field equation for $\phi$, \ref{bd2},  is basically a Poisson equation with an omnipresent source, $\Lambda$. Accordingly, even though $\omega$ is very large, perhaps we cannot rule out the possibility of existence of certain alternative boundary condition(s) that instead permits a strong  $\phi$  at very large scales.  Since $\Lambda$ couples directly to $\phi$, the first of \ref{bd2},  this could indicate a possible screening of the former. Under such circumstances, a cosmological event horizon, corresponding to the asymptotic de Sitter boundary condition may not exist, replaced by some suitable alternative asymptotic structure. What are these alternative boundary conditions? Can we have a regular stationary black hole solution in this scenario? 

One might try conceiving asymptotic(s) alternative to de Sitter or flat spacetimes, keeping only in mind the spacetime must be non-naked singular. However,  it seems that any such particular choice will be highly non-unique, being devoid of any symmetry argument in the asymptotic region. In order to tackle this difficulty, and to accommodate sufficient generality, we shall {\it not} at all impose any boundary condition as $r\to \infty$, and instead, we shall impose the same on the black hole event horizon. With this `initial condition', we wish to find out what happens at large scales, as follows. 
   
The recent discovery of gravity waves from the black hole mergers suggests that   the near horizon geometry matches exceedingly well with  the prediction of General Relativity~\cite{gravwaves}-\cite{Gerosa:2022fbk}. This, along with   the bound $|\omega| \gtrsim 40\,000$  suggests that the Brans-Dicke field must be weak at small scales such as a black hole. Hence we shall take the near black hole  horizon geometry to be the Schwarzschild-de Sitter at the leading order,
\be
ds^2\vert_{\rm BH} \to  -\left(1-\frac{2M}{r}-\frac{\Lambda r^2}{3}\right)dt^2+\left(1-\frac{2M}{r}-\frac{\Lambda r^2}{3}\right)^{-1}dr^2 + r^2\left(d\theta^2+\sin^2\theta d\varphi^2\right)
\label{bd7}
\ee
Clearly, this  necessitates that the field $\phi$ is close to unity and very slowly varying near the black hole horizon. We wish to show that this is indeed the case, as follows. The general solution of the second of \ref{bd2} with $T_{\mu\nu}=0$ in the background of \ref{bd7} is given by~\cite{Bhattacharya:2015iha},
\be
\phi(r )\vert_{\rm BH} \to C_2 + \frac{1}{\omega +3/2} \left[ \frac{C_1}{r_H} \ln \left(1-\frac{r_H}{r} \right)  +\left(1- \frac{C_1}{2 r_C}\right) \ln \left(1- \frac{r}{r_C} \right)+ \left(1+ \frac{C_1}{2 r_C}\right) \ln \left(1+ \frac{r}{r_C} \right) \right]
\label{bd8}
\ee
where $C_1,\,C_2$ are integration constants and $r_H$ is the black hole horizon radius (i.e., the smallest positive root of $1-2M/r-\Lambda r^2/3=0$) and $r_C = \sqrt{3/\Lambda}$. For the sake of computational simplicity, in the above derivation we have assumed that the black hole horizon is much small compared to $r_C$, owing to the observed current tiny value of $\Lambda\sim 10^{-52}{\rm m}^{-2}$, implying $M{\sqrt \Lambda}\ll 1$. Since as $\omega \to \infty$, we must have $\phi \to 1$, we set $C_2=1$.  The above solution is divergent on $r=r_H$. Thus a regular solution on it must correspond to $C_1=0$, giving  
\be
\phi(r \to r_H) \to 1 + \frac{1}{\omega +3/2} \ln \left(1- \frac{r^2}{r_C^2}\right)
\label{bd9}
\ee
Note that setting $\Lambda=0$ (i.e., $r_C\to \infty$) yields $\phi \to 1$, reproducing the no hair result of~\cite{Hawking}. Using now $r_C \sim 10^{26}{\rm m}$, it is easy to see that the dynamical part of the above solution is much small compared to unity. For example, even for a few billion solar mass black hole ($r_H \sim  10^{16}{\rm m}$), and $\omega \gtrsim 10^4$, it is at most ${\cal O}(10^{-24})$ and is further smaller for smaller black holes. Using now \ref{bd9},  \ref{bd4} into \ref{bd2}, it is also easy to find out the leading correction to the near horizon metric 
$$f(r\to r_H)= h^{-1}(r\to r_H) \to \left(1-\frac{2M}{r}-\frac{\Lambda r^2}{3}-\frac{\Lambda r^4}{5\omega r_C^2}\right) + {\cal O}\left(\frac{1}{\omega^2}\right)$$
For $r_H \sim  10^{16}{\rm m}$, the mass term is  ${\cal O}(1)$, the $\Lambda$ term is ${\cal O}(10^{-20})$, whereas the Brans-Dicke correction term is   ${\cal O}(10^{-45})$, or smaller.
Thus the backreaction on the spacetime near the black hole horizon due to $\phi(r)$ can safely be ignored.  We also note that any other choice of $C_1$ in \ref{bd8} leads to a curvature singularity on the black hole horizon~\cite{Bhattacharya:2015iha}. Thus \ref{bd9} is the unique  solution regular on the black hole and hence the boundary condition of \ref{bd7} is justified. 

Note that in our present scenario, if we construct now an integral equation like \ref{bd6}, although the boundary integral on the black hole horizon vanishes ($f(r_H),\,h^{-1}(r_H)=0$), the outer boundary integral at $r\to \infty$ does not, as $f$  and $h^{-1}$ is nowhere vanishing except on the black hole horizon, by our assumption. The simplest way to solve this problem seems to be to solve \ref{bd2}, \ref{bd4} numerically for general radial values, with the boundary conditions of \ref{bd7}, \ref{bd9}. However, we wish to do the same first in a much simpler and perhaps clearer analytic way, as follows.

Let us integrate \ref{bd5}  over the spacelike  hypersurces $(r,\theta,\phi)$ $\Sigma$ of \ref{bd4}, and convert the total divergence into surface integrals over $S^2$'s on the black hole horizon and on some larger radial coordinate, say $r_0$. As explained above, the surface integral on the black hole horizon vanishes  to yield,
  \be
4\pi r^2 \sqrt{\frac{f}{h}} \frac{d\phi}{dr}\Big \vert_{r=r_0}=  -\frac{4\Lambda}{2\omega +3} \int_{\Sigma}\sqrt{f} 
\label{bd10}
\ee
  The above equation shows that $\phi$ must be monotonic in $r$ in the entire domain $r_H \leq r <\infty$. For otherwise if it had any extremum at some $r=r_0$, we must have $\omega^{-1}=0$, since $\Lambda \neq 0$ by our assumption.  Also, since there is no horizon for $r>r_H $, $f(r)$ cannot be asymptotically vanishing and hence the integral on the right hand side of the above equation increases with increasing $r_0$. Thus since $\Lambda$ is positive, for $2\omega +3 >0$, $\phi$ is monotonically decreasing whereas it is monotonically increasing for $2\omega +3 <0$. Can $\phi(r)$ asymptote to some constant value? The answer is no, as this will simply correspond to asymptotic fall off in inverse power in $r$ for the dynamical part of $\phi$, leading to de Sitter geometry as dictated by the field equations \ref{bd2}. However, as we have stated earlier,  in the presence of an asymptotic de Sitter geometry, we must have $\omega^{-1}=0$~\cite{Bhattacharya:2015iha}.
  
Let us first consider $2\omega+3 >0$, i.e. effectively $\omega>0$, owing to the stringent solar system bound on it.  Note that the dynamical part of \ref{bd9} is negative, as $r_H <r_C$. Clearly, since $\phi(r)$ must be  monotonically decreasing, this dynamical part, as we move towards larger radial values, would decrease to further negative values, no matter what its explicit form is. Eventually thus the dynamical part will reach minus of unity, making $\phi(r)$ vanishing at that point. However,  the first term on the right hand side of \ref{bd3}, will diverge then. The second term contains a spacelike inner product since $\phi$ is explicitly independent of time. Thus \ref{bd3} is basically the sum of two positive definite quantities for $\phi \geq 0$. Hence no matter whether the second term diverges or not as $\phi \to 0$, the divergence of the first term sufficiently indicates the divergence of the Ricci scalar, $R \to \infty$, i.e. a naked curvature singularity. Thus we must have $\omega^{-1}=0$, in order to have a regular black hole spacetime, thereby reducing the theory to General Relativity, and hence we have the Schwarzschild-de Sitter to be the only solution.
  
Since $\phi(r)$ is positive on the horizon, \ref{bd9},  for $2\omega+3 <0$ (i.e., $\omega<0$ effectively) on the other hand, $\phi(r)$ is ever increasing to larger positive values with increasing $r$, and hence would eventually diverge as $r \to \infty$. We shall show in \ref{S3} that this also gives rise to a naked curvature singularity. However, even without this knowledge we wish to argue that having $\phi \to \infty$ is unphysical and unacceptable. This is because  since $\phi^{-1}$ should act as an effective Newton's `constant' in the Brans-Dicke theory, a divergent $\phi$ in the asymptotic region would make any matter field backreactionless there.  For example, let us imagine a static point mass $m$ at some $r=r_0$. We have its energy momentum tensor 
$$T_{\mu\nu} = m f(r) \delta_{\mu 0}\delta_{\nu 0} \delta^3(\vec{r}-\vec{r}_0)$$
substituting this into the first of \ref{bd2}, we may solve for corrections to $f$ and $h$ perturbatively due to $m$, assuming it is tiny. The corresponding potential due to $m$ will contain a $\phi(r_0)$ in the denominator. Since $\phi(r_0)$ diverges as $r_0 \to \infty$, the potential must vanish. This not only violates Mach's principle upon which the Brans-Dicke theory is based~\cite{Brans:1961sx, Brans:1962zz}, but also violates the most fundamental fact that any mass-energy distribution, no matter how tiny it is, must create its own gravity. To the best of our knowledge and understanding, the above scenario is physically unacceptable, thereby necessitating $\omega^{-1}=0$ in this case as well. Thus we may conclude that : there exists no regular or physically acceptable stationary black hole solution in the Brans-Dicke theory with a positive $\Lambda$, irrespective of whether asymptotically de Sitter boundary condition holds or not. We must have $\omega^{-1}=0$ and hence the solution corresponds to that of the General Relativity.  

The above result can easily be extended to rotating black hole spacetimes, by replacing the near horizon boundary condition, \ref{bd7}, with the Kerr-de Sitter spacetime. The near horizon solution for $\phi$ can be found in the form $\phi(r\to r_H) \to 1+\phi(r)+\phi(\theta)$, where $\phi(r)$ is analogous to \ref{bd9}, free of any singularity on the black hole horizon and $\phi(\theta) \to 0$ as we take the static limit. Similar argument as earlier then yields the same non-existence result.

In the next Section we shall further reinforce these non-existence features by explicit numerical computations.
\begin{figure}[h]
    \centering
    \includegraphics[scale=0.6]{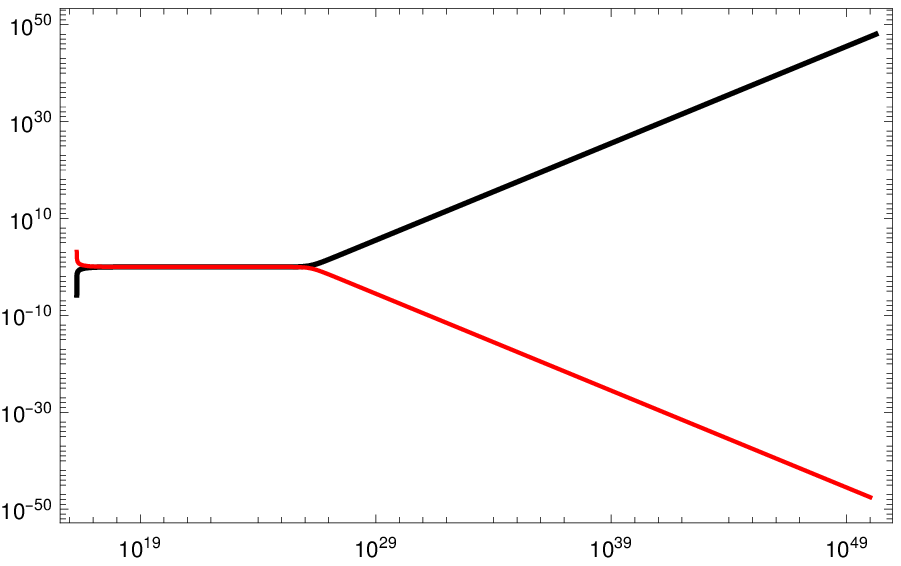}\hspace{2mm}
    \includegraphics[scale=0.6]{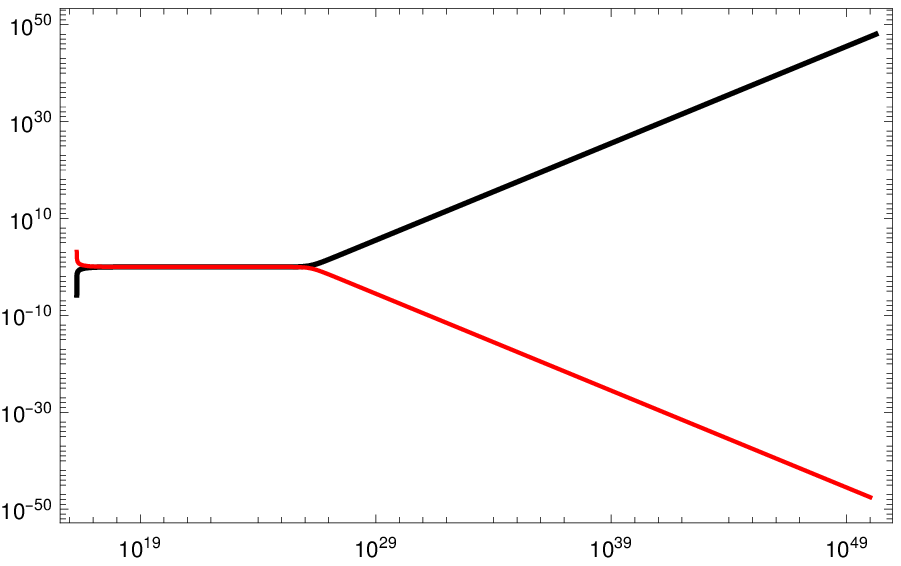}
    \caption{\small \it The variation of the metric functions $f(r)$ (black curve) and $h(r)$ (red curve) vs $r$. The first  plot corresponds to the Brans-Dicke parameter $\omega=40000$, whereas the right one corresponds to $\omega=-40000$. We have used the logarithmic scales on both the horizontal and vertical axes, in order to accommodate the large variations.}
    \label{f1}
\end{figure}
%

\subsection{Explicit numerical computations}\label{S3}

\begin{figure}
    \centering
    \includegraphics[scale=0.6]{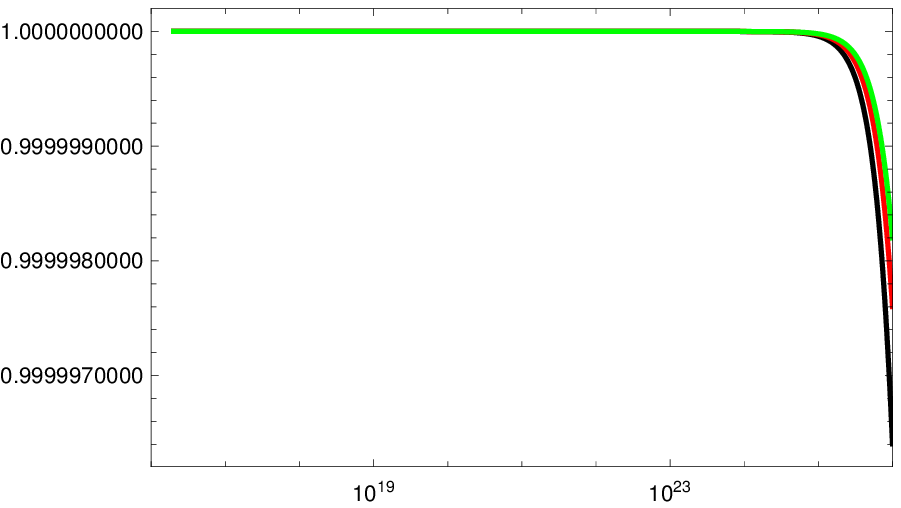}\hspace{2mm}
\includegraphics[scale=0.6]{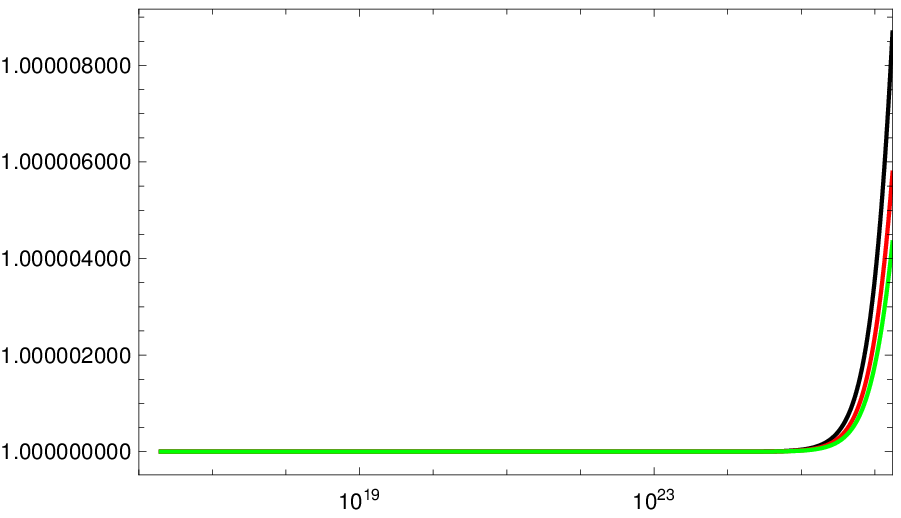}
    \caption{\small \it The Brans-Dicke field $\phi(r)$ vs $r$, for different values of the Brans-Dicke parameter $\omega$ showing its monotonic behaviour, as argued in \ref{S2}. The first set is for $\omega>0$, and the second is for $\omega<0$. The different colours correspond to : black ($\omega=\pm40000$), red ($\omega=\pm60000$) and green ($\omega=\pm80000$). }
        \label{f2}
\end{figure}

\begin{figure}[htbp]
    \centering
        \label{BD_field}
    \includegraphics[scale=0.7]{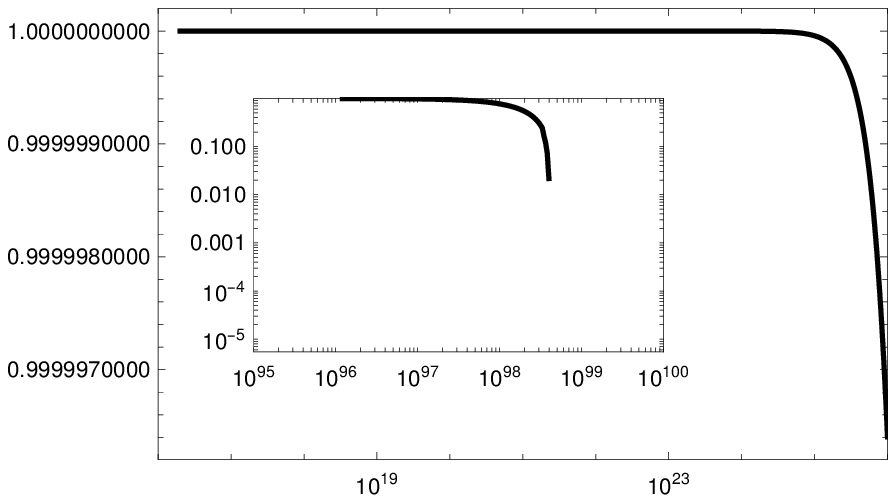}\hspace{2mm}
    \includegraphics[scale=0.6]{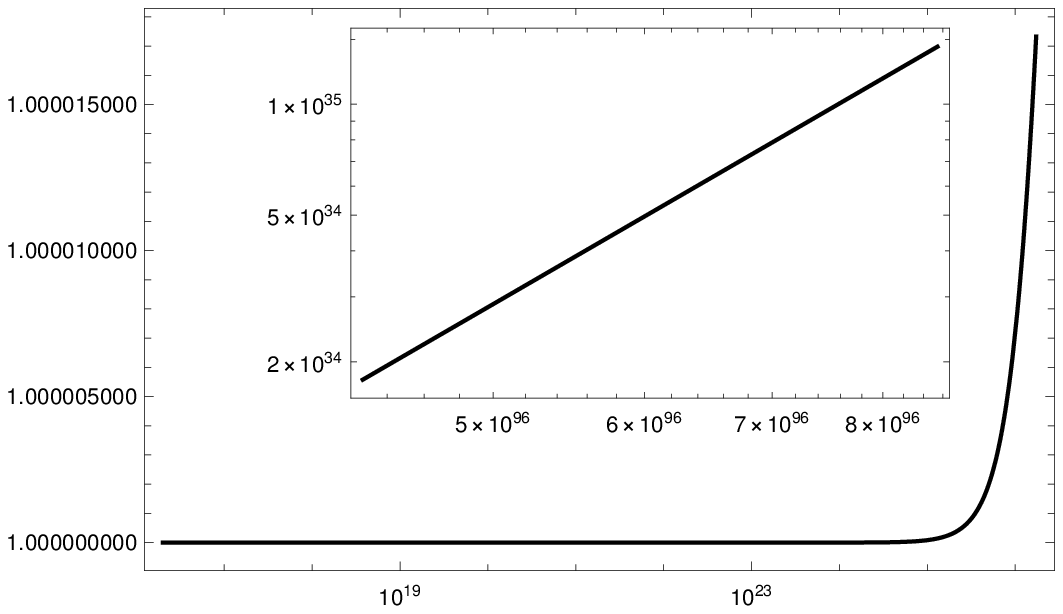}
  \caption{\small \it $\phi(r)$ vs $r$ with a magnified scale, depicting its pathological behaviour at large radial distances. The left and right plot respectively corresponds to $\omega=\pm 40000$. See main text for discussion. }
   \label{f3}
\end{figure}

\begin{figure}
    \centering
        \label{ricciscalar}
    \includegraphics[scale=0.46]{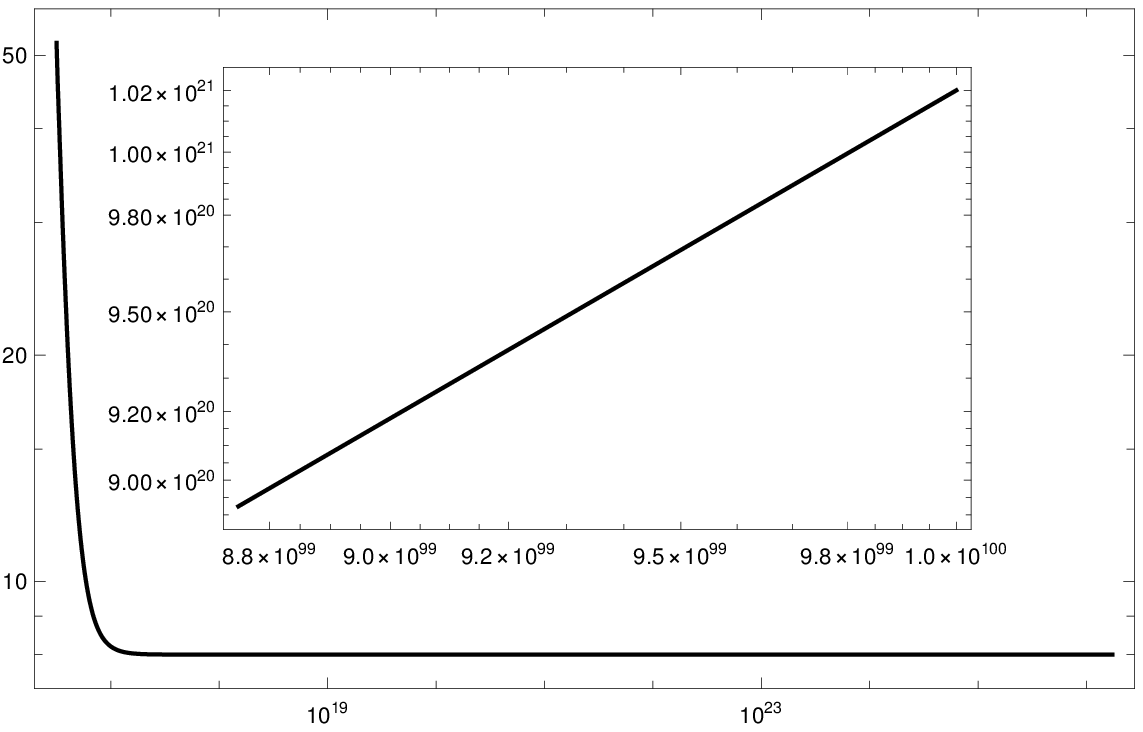}\hspace{2mm}
    \includegraphics[scale=0.60]{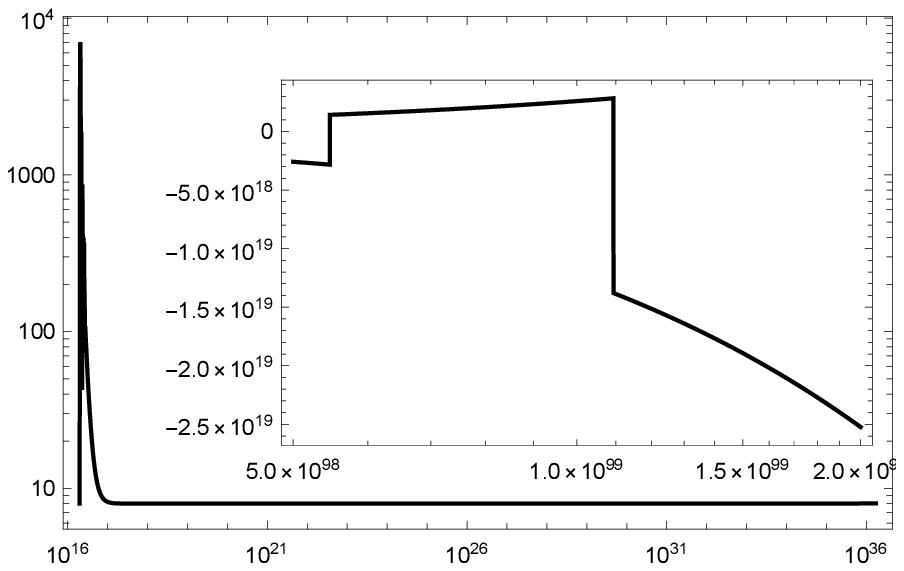}
    \caption{\small \it The variation of the Ricci scalar $R$, \ref{bd3},
 vs $r$. The left and right plot respectively corresponds to $\omega=\pm 40000$. We have used the logarithmic scales on both the horizontal and vertical axes. The plots in the inset are in the usual scale to show the asymptotic divergences in $R$. See main text for discussion.}
    \label{f4}
\end{figure}

Substituting \ref{bd4} into \ref{bd2}, we now solve for $f(r)$, $h(r)$ and $\phi(r)$ numerically using Mathematica with $T=0$, subject to the boundary conditions \ref{bd7}, \ref{bd9}. We take $2M \sim 10^{16}{\rm m}$ and $\Lambda \sim 10^{-52}{\rm m}^{-2}$. Using these we estimate the above three functions and their first derivatives' numerical values on the black hole event horizon, and further solve the coupled differential equations  for general $r > r_H$.     \ref{f1} shows the variation of the metric functions $f(r)$, $h(r)$ at large scales.    \ref{f2} shows the variation of $\phi(r)$, whereas  \ref{f3} depicts the same with a magnified resolution. Finally, using these results, we compute the Ricci scalar, \ref{bd3}, in \ref{f4}. We have taken $|\omega| \geq 40 000$, once again to be consistent with the recent observational evidence~\cite{Clifton:2011jh}.

First, \ref{f1} shows that subject to the boundary condition we have chosen on the black hole event horizon, there is indeed no cosmological event horizon (necessitating $f\to 0,\,h^{-1}\to 0$)  in the asymptotic region. \ref{f2} and \ref{f3} shows $\phi(r)$ indeed monotonically decreases and increases with $r$, respectively for $2\omega+3>0$ and $2\omega+3<0$, as was argued in the preceding Section. Also \ref{f3} shows that  $\phi(r)$ respectively passes through zero (diverges) for $2\omega+3>0$ ($2\omega+3<0$). Finally, \ref{f4} depicts the divergence of Ricci scalar for large radial values, thereby clearly proving that the spacetime we have obtained is naked singular and hence unacceptable.  Thus the numerical analysis presented here explicitly reconciles with the conclusion reached in the preceding Section.

%
\section{Non-existence of stationary star solutions}\label{S4}
We assume that the trace of the energy momentum-tensor constituting the star is less than or equal to zero. This is a reasonable assumption, indicating that the pressure  of the matter field is not `too large'~\cite{Wald:1984rg}. We also assume that the centre of the star is flat, owing to the fact that as we move closer to the centre, we have lesser and lesser matter fields to create gravity~\cite{Wald:1984rg}. There is no event horizon in this spacetime. Using the flatness at the centre,  we now integrate \ref{bd5} from $r=0$ up to the star surface to have for $d\phi/dr$, 
  \be
4\pi r^2 \sqrt{\frac{f}{h}} \frac{d\phi}{dr}\Big \vert_{r=R_0}=  -\frac{1}{2\omega +3} \int_{\Sigma}^{R_0}\sqrt{f} \left(4\Lambda -T \right)
\label{bd11}
\ee
where $R_0$ is radius of the star.  We next integrate \ref{bd5} from $r=R_0$ up to some $r=r_0$ outside the star and use \ref{bd11} into it to have
  \be
4\pi r^2 \sqrt{\frac{f}{h}} \frac{d\phi}{dr}\Big \vert_{r=r_0}=  -\frac{1}{2\omega +3} \int_{\Sigma, r=0}^{R_0}\sqrt{f} \left(4\Lambda -T \right)-\frac{4\Lambda}{2\omega +3} \int_{\Sigma, R_0}^{r_0}\sqrt{f} 
\label{bd12}
\ee
Thus as earlier, $d\phi/dr$ is monotonically decreasing (increasing) for $2\omega+3 >0$ ($2\omega+3 <0$).  Hence essentially the non-existence result of \ref{S2} holds for a star as well. 

\section{Conclusion}\label{S5}
We have discussed in this work the non-existence of regular stationary black hole and star solutions in the Brans-Dicke theory in the presence of a positive cosmological constant, $\Lambda$. It was shown earlier in~\cite{Bhattacharya:2015iha} that if a cosmological event horizon exists, we must have the inverse Brans-Dicke parameter $\omega^{-1}=0$, thereby reducing the theory to Einstein's General Relativity. As we have argued in \ref{S2}, even though it is reasonable to expect a cosmological event horizon in the asymptotic region owing to the repulsive effects of positive $\Lambda$, one cannot {\it a priori} rule out possible alternative boundary conditions where the Brans-Dicke field $\phi(r)$ is strong instead, and thereby screening $\Lambda$ at large scales. This corresponds to the fact that $\phi$ is sourced by an omnipresent $\Lambda$, and it couples directly with $\Lambda$, \ref{bd2}.  Note that once we discard the asymptotic de Sitter structure, we do not have any obvious symmetry argument to explicitly define an alternative one. Hence in order to make our analysis as generic as possible, we did not impose any boundary condition for large $r$, but did so instead on the black hole event horizon explicitly, inspired by the recent  gravity wave data coming from the black hole mergers~\cite{gravwaves}-\cite{Gerosa:2022fbk}. With this `initial condition', we showed in \ref{S2}, \ref{S3}, \ref{S4} that the existence of any non-naked singular stationary black hole as well as  star spacetimes with $\Lambda>0$ essentially necessitate, $\omega^{-1}=0$, thereby the theory reduces to Einstein's General Relativity in this scenario as well.   The black hole and the exterior of the star will then be described by the Kerr- or the Schwarzschild-de Sitter spacetime. 

We would like to emphasise the stark contrast that while the standard no hair theorems only talk about  the field configurations, e.g.~\cite{Hawking},  we obtain the parameter characterising  a theory is getting constrained, not only for black holes but also for stars. Note also that the non-trivial effects due to a positive $\Lambda$ reported in~\cite{Bhattacharya:2015iha} (also in e.g.~\cite{Bhattacharya:2007ap,  Ashtekar:2015lla}), was chiefly related to the exotic boundary effects due to the cosmological event horizon. This is contrary to our present case, as no such horizon was assumed to be present and hence no boundary effect was possible here. Since the Brans-Dicke theory is considered to be the prototype of the scalar-tensor class of alternative gravity theories, we believe the result we have found to be interesting and important in its own right. 

It seems to be an interesting task to check the cosmological anisotropy dissipation/no hair theorem of~\cite{Wald} in the context of the Brans-Dicke theory with a positive $\Lambda$. We hope to return to this issue in a future work.

\section*{Acknowledgement}
 The research of MSA
is supported by the National Postdoctoral Fellowship
of the Science and Engineering Research Board (SERB),
Department of Science and Technology (DST),
Government of India, File No. PDF/2021/00349. SB acknowledges partial support from the startup grant (S-3/122/22) of Jadavpur University, Kolkata, India.

  

\begin{thebibliography}{}
 
 \bibitem{Bekenstein:1971hc}
J.~D.~Bekenstein,
{\it Nonexistence of baryon number for static black holes},
Phys. Rev. D \textbf{5}, 1239-1246 (1972)
 
 \bibitem{Bekenstein:1998aw}
J.~D.~Bekenstein,
{\it Black holes: Classical properties, thermodynamics and heuristic quantization},
[arXiv:gr-qc/9808028 [gr-qc]].

\bibitem{Herdeiro:2015waa}
C.~A.~R.~Herdeiro and E.~Radu,
{\it Asymptotically flat black holes with scalar hair: a review},
Int. J. Mod. Phys. D \textbf{24}, no.09, 1542014 (2015)
[arXiv:1504.08209 [gr-qc]].
 
 \bibitem{Weinberg:2008zzc}
S.~Weinberg, {\it Cosmology}, Oxford Univ. Press (2009).

\bibitem{Gibbons:1977mu}
G.~W.~Gibbons and S.~W.~Hawking,
{\it Cosmological Event Horizons, Thermodynamics, and Particle Creation},
Phys. Rev. D \textbf{15}, 2738-2751 (1977)

 
 \bibitem{Chambers:1994sz}
C.~M.~Chambers and I.~G.~Moss,
{\it A Cosmological no hair theorem},
Phys. Rev. Lett. \textbf{73}, 617-620 (1994)
[arXiv:gr-qc/9406036 [gr-qc]].
 
 \bibitem{Bhattacharya:2007ap}
S.~Bhattacharya and A.~Lahiri,
{\it Black-hole no-hair theorems for a positive cosmological constant},
Phys. Rev. Lett. \textbf{99}, 201101 (2007)
[arXiv:gr-qc/0702006 [gr-qc]].


\bibitem{Bhattacharya:2011dq}
S.~Bhattacharya and A.~Lahiri,
{\it No hair theorems for stationary axisymmetric black holes},
Phys. Rev. D \textbf{83}, 124017 (2011)
[arXiv:1102.0053 [gr-qc]].

\bibitem{Bhattacharya:2013hvm}
S.~Bhattacharya and H.~Maeda,
{\it Can a black hole with conformal scalar hair rotate?},
Phys. Rev. D \textbf{89}, no.8, 087501 (2014)
[arXiv:1311.0087 [gr-qc]].


\bibitem{Clifton:2011jh}
T.~Clifton, P.~G.~Ferreira, A.~Padilla and C.~Skordis,
{\it Modified Gravity and Cosmology},
Phys. Rept. \textbf{513}, 1-189 (2012)
[arXiv:1106.2476 [astro-ph.CO]].
 
 \bibitem{Brans:1961sx}
C.~Brans and R.~H.~Dicke,
{\it Mach's principle and a relativistic theory of gravitation},
Phys. Rev.\textbf{124}, 925-935 (1961).

\bibitem{Brans:1962zz}
C.~H.~Brans,
{\it Mach's Principle and a Relativistic Theory of Gravitation. II},
Phys. Rev.\textbf{125}, 2194-2201 (1962).


\bibitem{Hawking}
S. W. Hawking, {\it Black holes in the Brans-Dicke theory of gravitation}, Comm. Math. Phys. \textbf{25} 167-171 (1972).

\bibitem{Sotiriou:2011dz}
T.~P.~Sotiriou and V.~Faraoni,
{\it Black holes in scalar-tensor gravity},
Phys. Rev. Lett. \textbf{108}, 081103 (2012)
[arXiv:1109.6324 [gr-qc]].

\bibitem{Hui:2012qt}
L.~Hui and A.~Nicolis,
{\it No-Hair Theorem for the Galileon},
Phys. Rev. Lett. \textbf{110}, 241104 (2013)
[arXiv:1202.1296 [hep-th]].

\bibitem{Sotiriou:2013qea}
T.~P.~Sotiriou and S.~Y.~Zhou,
{\it Black hole hair in generalized scalar-tensor gravity},
Phys. Rev. Lett. \textbf{112}, 251102 (2014)
[arXiv:1312.3622 [gr-qc]].

\bibitem{Babichev:2015rva}
E.~Babichev, C.~Charmousis and M.~Hassaine,
{\it Charged Galileon black holes},
JCAP \textbf{05}, 031 (2015)
[arXiv:1503.02545 [gr-qc]].

\bibitem{Graham:2014mda}
A.~A.~H.~Graham and R.~Jha,
{\it Nonexistence of black holes with noncanonical scalar fields},
Phys. Rev. D \textbf{89}, no.8, 084056 (2014)
[erratum: Phys. Rev. D \textbf{92}, no.6, 069901 (2015)]
[arXiv:1401.8203 [gr-qc]].

\bibitem{Bhattacharya:2015iha}
S.~Bhattacharya, K.~F.~Dialektopoulos, A.~E.~Romano and T.~N.~Tomaras,
{\it Brans-Dicke Theory with $\Lambda>0$: Black Holes and Large Scale Structures},
Phys. Rev. Lett.\textbf{115}, no.18, 181104 (2015)
[arXiv:1505.02375 [gr-qc]].

\bibitem{Bhattacharya:2016vur}
S.~Bhattacharya, K.~F.~Dialektopoulos, A.~E.~Romano, C.~Skordis and T.~N.~Tomaras,
{\it The maximum sizes of large scale structures in alternative theories of gravity},
JCAP\textbf{07}, 018 (2017)
[arXiv:1611.05055 [astro-ph.CO]].

\bibitem{Maeda:2011jj}
H.~Maeda and G.~Giribet,
{\it Lifshitz black holes in Brans-Dicke theory},
JHEP \textbf{11}, 015 (2011)
[arXiv:1105.1331 [gr-qc]].

\bibitem{Faraoni:2021nhi}
V.~Faraoni, A.~Giusti and B.~H.~Fahim,
{\it Spherical inhomogeneous solutions of Einstein and scalar\textendash{}tensor gravity: A map of the land},
Phys. Rept. \textbf{925}, 1-58 (2021)
[arXiv:2101.00266 [gr-qc]].

\bibitem{Nojiri:2020blr}
S.~Nojiri, S.~D.~Odintsov and V.~Faraoni,
{\it Searching for dynamical black holes in various theories of gravity},
Phys. Rev. D \textbf{103}, no.4, 044055 (2021)
[arXiv:2010.11790 [gr-qc]].

\bibitem{Papantonopoulos:2019ugr}
E.~Papantonopoulos and C.~Vlachos,
{\it Wormhole solutions in modified Brans-Dicke theory},
Phys. Rev. D \textbf{101}, no.6, 064025 (2020)
[arXiv:1912.04005 [gr-qc]].

\bibitem{Ranjbar:2011rf}
A.~Ranjbar, H.~R.~Sepangi and S.~Shahidi,
{\it Asymptotically Lifshitz Brane-World Black Holes},
Annals Phys. \textbf{327}, 3170 (2012)
[arXiv:1108.4562 [hep-th]].

\bibitem{SantaVelez:2019woi}
C.~Santa V\'elez and A.~Enea Romano,
{\it Constraining gravity theories with the gravitational stability mass},
JCAP \textbf{06}, 022 (2020)
[arXiv:1905.07620 [gr-qc]].

\bibitem{Faraoni:2018nql}
V.~Faraoni and J.~C\^ot\'e,
{\it Two new approaches to the anomalous limit of Brans-Dicke theory to Einstein gravity},
Phys. Rev. D \textbf{99}, no.6, 064013 (2019)
[arXiv:1811.01728 [gr-qc]].

\bibitem{Faraoni:2017ock}
V.~Faraoni,
{\it Jordan frame no-hair for spherical scalar-tensor black holes},
Phys. Rev. D \textbf{95}, no.12, 124013 (2017)
[arXiv:1705.07134 [gr-qc]].

\bibitem{Scheel:1994yr}
M.~A.~Scheel, S.~L.~Shapiro and S.~A.~Teukolsky,
{\it Collapse to black holes in Brans-Dicke theory. 1. Horizon boundary conditions for dynamical space-times},
Phys. Rev. D \textbf{51}, 4208-4235 (1995), 
[arXiv:gr-qc/9411025 [gr-qc]].

\bibitem{Kang:1996rj}
G.~Kang,
{\it On black hole area in Brans-Dicke theory},
Phys. Rev. D \textbf{54}, 7483-7489 (1996)
[arXiv:gr-qc/9606020 [gr-qc]].

\bibitem{Bhattacharya:2016lup}
S.~Bhattacharya,
{\it Rotating Killing horizons in generic $F(R)$ gravity theories},
Gen. Rel. Grav. \textbf{48}, no.10, 128 (2016)
[arXiv:1602.04306 [gr-qc]].

\bibitem{Nojiri:2020blr}
S.~Nojiri, S.~D.~Odintsov and V.~Faraoni,
{\it Searching for dynamical black holes in various theories of gravity},
Phys. Rev. D \textbf{103}, no.4, 044055 (2021)
[arXiv:2010.11790 [gr-qc]].



\bibitem{gravwaves}
B. P. Abbott \textit{et al.}, {\it Observation of Gravitational Waves from a Binary Black Hole Merger}, Phys. Rev. Lett. \textbf{116}, 061102 (2016).


\bibitem{LIGOScientific:2016lio}
B.~P.~Abbott \textit{et al.} [LIGO Scientific and Virgo],
{\it Tests of general relativity with GW150914},
Phys. Rev. Lett. \textbf{116}, 221101 (2016)
[erratum: Phys. Rev. Lett. \textbf{121}, 129902 (2018)]
[arXiv:1602.03841 [gr-qc]].

\bibitem{LIGOScientific:2018dkp}
B.~P.~Abbott \textit{et al.} [LIGO Scientific and Virgo],
{\it Tests of General Relativity with GW170817},
Phys. Rev. Lett. \textbf{123}, 011102 (2019)
[arXiv:1811.00364 [gr-qc]].

\bibitem{LIGOScientific:2019fpa}
B.~P.~Abbott \textit{et al.} [LIGO Scientific and Virgo],
{\it Tests of General Relativity with the Binary Black Hole Signals from the LIGO-Virgo Catalog GWTC-1},
Phys. Rev. D \textbf{100}, 104036 (2019)
[arXiv:1903.04467 [gr-qc]].

\bibitem{Prasad:2020xgr}
V.~Prasad, A.~Gupta, S.~Bose, B.~Krishnan and E.~Schnetter,
{\it News from horizons in binary black hole mergers},
Phys. Rev. Lett. \textbf{125}, 121101 (2020)
[arXiv:2003.06215 [gr-qc]]

  


\bibitem{Isi:2020tac}
M.~Isi, W.~M.~Farr, M.~Giesler, M.~A.~Scheel and S.~A.~Teukolsky,
{\it Testing the Black-Hole Area Law with GW150914},
Phys. Rev. Lett. \textbf{127}, no.1, 011103 (2021)
[arXiv:2012.04486 [gr-qc]].


\bibitem{Gerosa:2022fbk}
D.~Gerosa, C.~M.~Fabbri and U.~Sperhake,
{\it The irreducible mass and the horizon area of LIGO\textquoteright{}s black holes},
Class. Quant. Grav. \textbf{39}, 175008 (2022)
[arXiv:2202.08848 [gr-qc]].


\bibitem{Wald:1984rg}
R.~M.~Wald, {\it General Relativity},
Chicago Univ. Press (USA), 1984.

\bibitem{Ashtekar:2015lla}
A.~Ashtekar, B.~Bonga and A.~Kesavan,
{\it Asymptotics with a positive cosmological constant. II. Linear fields on de Sitter spacetime},
Phys. Rev. D \textbf{92}, no.4, 044011 (2015)
[arXiv:1506.06152 [gr-qc]].



\bibitem{Wald}
R. M. Wald, {\it Asymptotic behavior of homogeneous cosmological models in the presence of a positive cosmological constant}, Phys. Rev. D \textbf{28}, 2118 (1983).


\end{thebibliography}
\end{document}